# PRINCIPAL COMPONENT ANALYSIS AND AUTOMATIC RELEVANCE DETERMINATION IN DAMAGE IDENTIFICATION

**L. Mdlazi, T. Marwala, C.J. Stander, C. Scheffer** and **P.S. Heyns**

Dynamic Systems Group
Department of Mechanical and Aeronautical Engineering
University of Pretoria
Pretoria, 0002, South Africa

**Abstract**

This paper compares two neural network input selection schemes, the Principal Component Analysis (PCA) and the Automatic Relevance Determination (ARD) based on MacKay's evidence framework. The PCA takes all the input data and projects it onto a lower dimension space, thereby reducing the dimension of the input space. This input reduction method often results with parameters that have significant influence on the dynamics of the data being diluted by those that do not influence the dynamics of the data. The ARD selects the most relevant input parameters and discards those that do not contribute significantly to the dynamics of the data being modelled. The ARD sometimes results with important input parameters being discarded thereby compromising the dynamics of the data. The PCA and ARD methods are implemented together with a Multi-Layer-Perceptron (MLP) network for fault identification in structures and the performance of the two methods is assessed. It is observed that ARD and PCA give similar accuracy levels when used as input-selection schemes. Therefore, the choice of input-selection scheme is dependent on the nature of the data being processed.

**1. Introduction**

Condition monitoring of critical machinery plays a very important role in modern day maintenance. A novel approach is using Artificial Neural Networks (ANN) based vibration analysis. Condition monitoring allows maintenance personnel to maximise machine availability by reducing machine down time and associated losses. With neural networks, researchers and practitioners are in many instances faced with the problem of dealing with large data inputs. Large input spaces are computationally expensive and some of the data in the input space may be totally unrelated to the output or target values. The inclusion of redundant data may sometimes reduce the network's classification accuracy.

Engineering judgement and the Principal Component Analysis (PCA) [1] have been used to select relevant inputs and reduce the input space to the neural network respectively. PCA is defined as a mathematical procedure that transforms a number of (possibly) correlated variables into a (smaller) number of uncorrelated variables called principal components. Another technique for reducing the input space called the Automatic Relevance Determination (ARD) was developed by Mackay in 1994 [2]. This method was developed further by Neal in 1996 [3].

The ARD determines the relevance of each input parameter. Informally, the aim of ARD is to discover which hidden variables are relevant in explaining the dynamics of the system of interest. The irrelevant variables are then pruned away. The ARD can be implemented as a form of Bayesian structure learning where a prior Gaussian distribution is placed on the weights, favouring small magnitudes. The essence of ARD is that each input unit has its own prior variance parameter. Small variance suggests that all weights leaving the unit will be small, so the unit will have little influence on subsequent values. A large variance indicates that the unit is important in explaining the data [2].

The aim of this study is to introduce ARD to the vibration community and compare ARD to PCA focusing on the practical implementation issues of the two input-selection schemes. Data from a cylinder experiment [3] and a gear vibration test rig [6] are used as inputs to the neural networks. The data are pre-processed using the ARD and PCA. The pre-processed data are used to train a Multi-Layer Perceptron (MLP) neural network. The trained neural networks are used for simulation and the network classifications for the two methods are assessed.

## 2. Experimental Setups

### 2.1 Cylinder experiment

Two data sets from different experiments are used as inputs to the neural network. The first data set is the modal properties from a cylinder experiment [3]. In this experiment, an impulse hammer test is performed on each of 20 steel seam-welded cylindrical shells (1.75 ±0.02 mm thickness, 101.86 ± 0.29 mm diameter and 101.4 ±0.2 mm). The cylinders are placed on a 'bubble wrap', to simulate a free-free environment (see Fig.1).

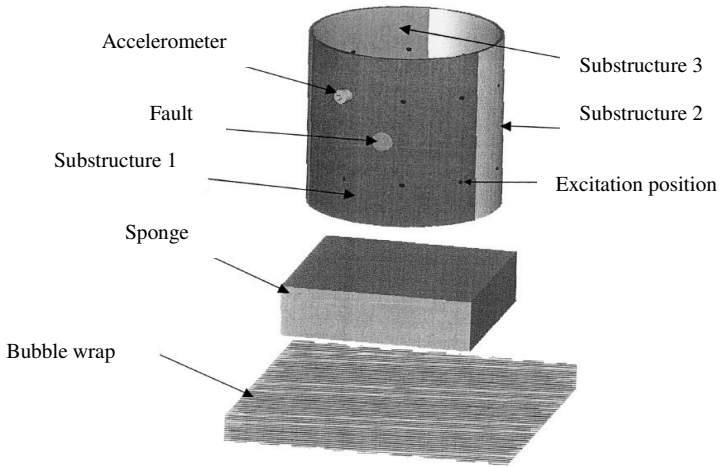

Fig. 1. Illustration of cylindrical shell showing the impulse position, accelerometer, substructure, fault position and support. (see [4])

A sponge is inserted inside the cylinder to control boundary conditions by rotating it each time a measurement is taken. The top impulse positions are located 25mm from the top edge of the cylinder and the bottom impulse positions are located 25mm from the bottom edge of the cylinder. The angle between two adjacent impulse positions is 36°. The holes are located at the centre of the substructure with radius of 10-15 mm. The holes are excited using a modal hammer with a sensitivity of 4pC/N; the mass of the head is 6.6g, and the cut off frequency is 3.64 kHz. The response is measured using an accelerometer with a sensitivity of 2.6pC/ms$^{-2}$, which has a mass of 19.8 g. Conventional signal processing procedures are applied to convert the time domain impulse history and response data into the frequency domain. The excitation and response data in the time domain are utilised to calculate the Frequency Response Functions (FRFs) and modal analysis is used to extract modal properties [3] from the FRFs.

Each cylinder is divided into three substructures, and holes of 10-15 mm in diameter are drilled on each substructure (see Fig.1). For one cylinder, there are no faults present. This is termed a zero-fault scenario. This type of fault is given the identity [0 0 0], indicating that there are no faults in any of the substructures. The second type of fault is a one-fault scenario, where a hole may be located in one of the substructures, 1, 2 or 3 respectively. Three possible one-fault scenarios are [1 0 0], [0 1 0] and [0 0 1] indicating the presence of one hole in the substructures 1, 2 or 3 respectively. The third type of fault is the two-fault scenario, where one hole is located in two of the three substructures. Three possible two-fault scenarios are [1 0 1], [1 1 0] and [0 1 1]. The final type of fault is the three-fault scenario, where a fault is located in all three substructures, and the identity of this fault is [1 1 1]. There are in total eight different types of fault cases considered (including [0 0 0]). For each damage case, data are obtained by measuring the acceleration at a fixed position and roving the impulse position. Each cylinder has four damage scenarios and the total number of damage cases collected is 264. The structure is vibrated in three different positions [3].

### 2.2 Gear vibration setup

The second data set is from a gear vibration test set-up. The test set-up consisted of a single-stage spur gearbox, driven by a 5 hp Dodge Silicon Controlled Rectifier motor. A 5.5 kVA Mecc alte spa three-phase alternator was used for applying the load. Figure 2 illustrates the test rig. The gears were manufactured in accordance with DIN3961, Quality 3 and had a load rating of 20Nm.

The Alternating Current (AC) generated by the alternator was rectified and dissipated over a large resistive load, which was kept constant during the tests. The Direct Current (DC) fields of the alternator were powered by an external DC supply in order to control the load that was applied to the gears. A single-phase voltage feedback from the alternator was used in conjunction with Proportional Integral Compensation to regulate the torque applied by the alternator. Tyre couplings were fitted between the electrical machines and the gearbox so that the backlash in the system would be restricted to the gears.

The average shaft speed during experimentation was 13 Hz. A synchronising pulse was measured by means of a proximity switch on the key of the shaft. Acceleration measurements were taken in the vertical direction with a 500 mV/g PCB Integrated Circuit Piezoelectric industrial accelerometer and a Siglab model 20-42 signal analyser.

The initial vibration measurements were taken without any induced damage or 0% damage. Then face wear was induced on one of the gear teeth by artificially removing material from the gear face. In addition, a crack was induced on the opposite side of the gear. The damage and fault identification details are presented in Table 1. The fault severity conditions are expressed as the fraction of the root crack length over the 4 mm tooth thickness.

TABLE 1 Induced damage specifications and fault identification.

| Damage | Fault severity 0% | Fault severity 25% | Fault severity 50% |
|---|---|---|---|
| Material removed from face | 0 mm nominally | 0.15 mm nominally | 0.3 mm nominally |
| Crack length | 0 mm | 1 mm | 2 mm |
| Fault identity | 0 | 0.5 | 1 |

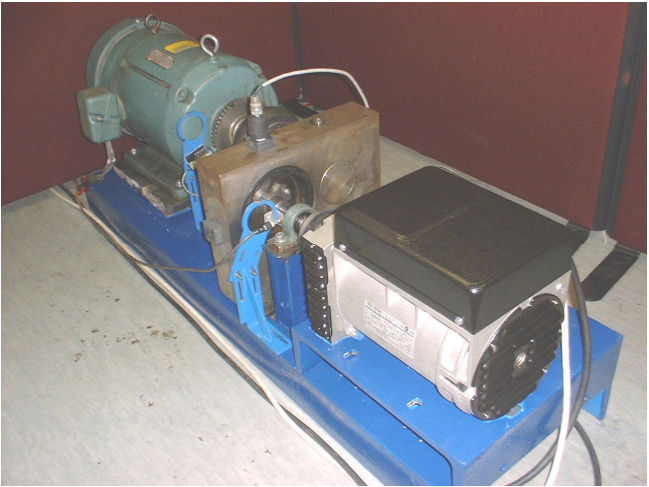

Fig. 2. Gear test rig set-up

## 3. Processing of the cylinder experiment data

### 3.1 Pre-processing

This section deals with the processing of the cylinder experiment data. To select the modal properties as input data and the following procedure was followed. First, the Statistical Overlap Factor (SOF) is calculated to investigate the presence of various noise levels in the input space. The SOF between two distributions is defined as the ratio of the distance between the averages of the two distributions, to the mean of the two standard deviations. The SOF may be written mathematically as follows:

$$SOF = \left| \frac{\overline{x_1} - \overline{x_2}}{(\sigma_1 + \sigma_2)/2} \right| \quad \text{(Eq.1)}$$

Here $\overline{x_1}$ and $\overline{x_2}$ are the means of distributions and $\sigma_1$ and $\sigma_2$ are their respective standard deviations.

The higher the SOF, the better the degree of separation between the two distributions. The SOF was implemented in the following five steps:

1. Find the mean and standard deviation of the modal properties at each index for data from undamaged and damaged cylinders.
2. Calculate the difference between the means of data from undamaged and damaged cylinders at each index.
3. Calculate the average of the standard deviations from undamaged and damaged cylinders while keeping track of the indices.
4. Calculate the ratio between the mean-difference in step (2) and the average-standard deviations difference in step (3) at each index.
5. From these ratios, select 50 indices with the highest ratios and use the corresponding data as input to ARD and PCA.

### 3.2 Implementing PCA

PCA is employed to reduce the 50 chosen inputs using statistical overlap factor to 10, 7, 5, and 3 independent variables respectively. The PCA is implemented by calculating the covariance matrix of the input data. The eigenvalues and eigenvectors of the covariance matrix are then calculated. The ten eigenvectors corresponding to the 10 largest eigenvalues are retained. The input data are then projected onto the corresponding eigenvectors. The new input data have a dimension of 10. This procedure is repeated to reduce the input space to 7, 5 and 3 respectively. The properties corresponding to the chosen indices are used as inputs for training four different networks.

### 3.3 Implementing ARD

ARD is implemented using a multi-layer perceptron. The output from the SOF is used as the input data to ARD. The target variables are the 264 fault identities from the experiment. The prior over weights are given by the ARD Gaussian prior with a separate hyper-parameter for the group of weights associated with each input. The network is trained by error minimization using the scaled conjugate gradient function (SCG) [4,5]. There are two cycles of training, and at the end of each cycle the hyper-parameters are re-estimated. The first 50 hyper parameters correspond to the inputs to hidden unit weights. The remaining hyper-parameters correspond to the hidden unit biases, second layer weights and output unit biases, respectively. Since each hyper-parameter corresponds to an inverse variance, the posterior variance for the weights associated with relevant data is large, and the weights associated with irrelevant data is small. The network gives greatest emphasis to relevant data and least emphasis to irrelevant data.

The weights assigned by ARD are sorted and the 10 input vectors corresponding to the 10 largest weights are retained as input data and the rest discarded. This results in a new input space of 10. This procedure is repeated to reduce the input space to 7, 5 and 3, respectively. The new input spaces are used as inputs for training four different neural networks.

### 3.4 Network classification

Four types of classifications are used. These are: The false negative case where the network falsely indicates the absence of faults. This case is the most undesirable because it falsely indicates that the situation is under control, which may have catastrophic consequences to critical components. The second type of classification is the false positive case where the network falsely indicates the presence of faults. This case is the second least desired because its consequences are economical rather than catastrophic. The other types of classification are the true positives and true negatives, which are the correct classifications. Classification results for ARD and PCA processing on the cylinder experiment are tabulated in Table 2 and plotted in Fig 3.

Table 2. Classification results for ARD and PCA pre-processing on the cylinder experiment data.

| Number of inputs | **PCA** Classification | **ARD** Classification |
|---|---|---|
| 3 | 61.36 | 51.15 |
| 5 | 80.68 | 68.18 |
| 7 | 90.53 | 83.33 |
| 10 | 93.18 | 91.29 |

From Fig.3 it can be seen that PCA performs slightly better than ARD. The best network classification is achieved with an input dimension of 10 for both methods. With PCA an average network classification of 93.18 % is obtained, while ARD gives an average network classification of 91.29 % after training and simulating five different networks for each method

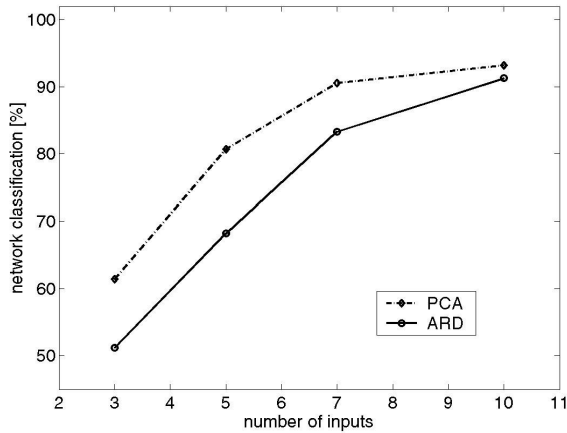

Fig 3. Network classification results for the Cylinder experiment.

## 4. Processing the gear vibration data

### 4.1 Pre-processing

This section deals with the processing of gear vibration data. The data has 828 examples. Each example represents 1 gear revolution. The first 276 examples are time domain vibration data from 276 revolutions of a gear with no damage. The second 276 examples are time domain vibration data from a gear with 50% damaged and the last 276 examples are time domain vibration data from 276 revolutions from a gear with 100 % damage. Each example has 1024 points. This results in a 828 x 1024 input matrix (input space). The data are pre-processed in four different ways before using ARD and PCA:

1. The first method is leaving the data in the time domain, reducing the number of examples per damage case to 100 and resampling the data at a lower frequency to reduce the number of points per revolution to 256. This results in a (300 x 256) input matrix. This input matrix is then used as the input to PCA and ARD.
2. The second method is leaving the data in the time domain, reducing the number of examples per damage case to 100 and resampling the data at a lower frequency to reduce the number of points per revolution to 64. This results in a (300 x 64 input matrix).
3. The third method is transforming the data generated in (1) to the frequency domain. This results in a 300 x 256 frequency domain input matrix. This input matrix is then used as the input to PCA and ARD.
4. The fourth method is leaving the data in the time domain, reducing the number of examples per damage case to 100 examples and applying the following feature extraction methods to reduce the number of points per fault case. These features include: mean, rms, crest factor, variance, skewness, kurtosis, AR coefficients, MA coefficients and ARMA coefficients. This results in a 300 x 62 input matrix. [7,8]

### 4.2 Implementing PCA

PCA is employed to reduce the input data from the four pre-processing techniques to 10, 7, 5 and 3 independent variables respectively. The PCA is implemented by calculating the covariance matrix of the input data. The eigenvalues and eigenvectors of the covariance matrix are then calculated. The ten eigenvectors corresponding to the 10 largest eigenvalues are retained. The input data from the pre-processing procedures are then projected onto the corresponding eigenvectors. The new input data have a dimension of 10. This procedure is repeated to reduce the input space to 7, 5 and 3, respectively. The properties corresponding to the chosen indices are used as inputs for training four different networks for each processing technique.

### 4.3 Implementing ARD

ARD is implemented using a multi-layer perceptron. The input data is the output from the four pre-processing techniques. The single target variables are the 300 fault identities from the gear experiment. The prior over weights are given by the ARD Gaussian prior with a separate hyper-parameter for all the weights associated with each input.

The network is trained by error minimization using scaled conjugate gradient function (SCG). There are two cycles of training, and at the end of each cycle the hyper-parameters are re-estimated. The first 300 hyper-parameters correspond to the inputs to the hidden unit weights. The remaining hyper-parameters correspond to the hidden unit biases, second layer weights and output unit biases, respectively. Since each hyper-parameter corresponds to an inverse variance, the posterior variance for weights associated with relevant data is large, and the weights associated with irrelevant data is small. The network gives greatest emphasis to relevant data and least emphasis to irrelevant data.

The weights assigned by ARD are sorted and the ten input vectors corresponding to the 10 largest weights are retained. This input data are used as the new input space discarding the rest of the data. This results in a new input space of ten. This procedure is repeated to reduce the input space to 7, 5 and 3 respectively.

The new input spaces are used as inputs for training 4 different networks for each of the pre-processing techniques.

### 4.4 Network classification results

Classification results for ARD and PCA for the different pre-processing methods on the gear vibration data are tabulated in Tables 3 to 6 and are plotted in Figs. 4 to 7.

#### 4.4.1 Time domain input data with 256 points as input

From Fig.4 it can be seen that PCA performs better than ARD. PCA gives a constant network classification of about 97.5 % for all four input sizes. For ARD a network classification of 92.3% is achieved at 10 inputs. There is a gradual increase in the classification as the input dimension is increased. This is because some of the discarded inputs contain valuable information in this data set. PCA yields a constant performance irrespective of the input dimension. This is because none of the data is discarded in PCA, but the data is simply transformed into a smaller input dimension.

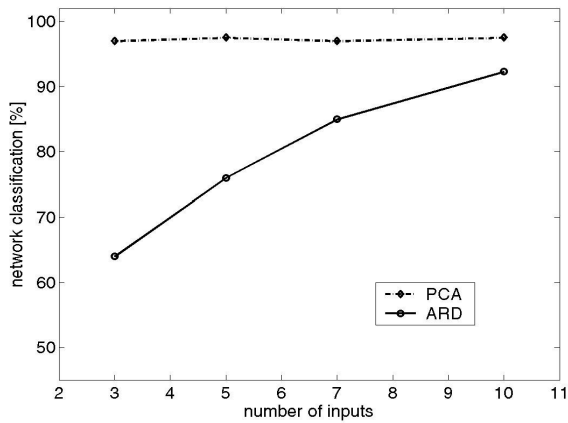

Fig. 4. Network classification results for the time domain gear vibration data with 256 points.

Table 3. Network classification results for ARD and PCA on time domain vibration data with 256 points.

| Number of inputs | PCA Classification | ARD Classification |
|---|---|---|
| 3 | 97.0 | 64.0 |
| 5 | 97.5 | 76.0 |
| 7 | 97.0 | 85.0 |
| 10 | 97.5 | 92.3 |

#### 4.4.2 Time domain input data with 64 points as input

From Fig.5 it can be seen that PCA performs better than ARD. Both PCA and ARD give the best network classification at an input dimension of 10. PCA gives a network classification of 61% while ARD gives a network classification of 58%. The poor performance is an indication that the sampling frequency that is used for this data set is too low. Most of the relevant information is contained in the higher frequencies.

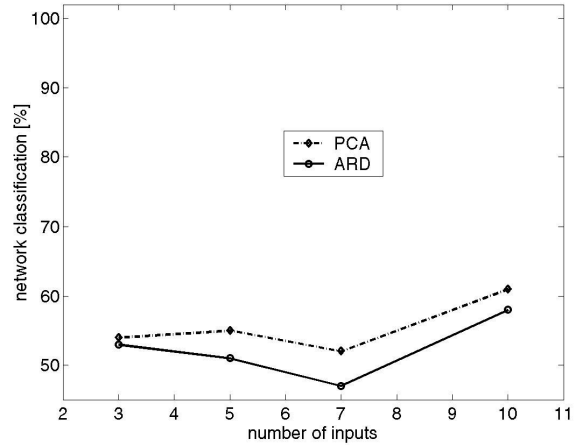

Fig. 5. Network classification results for the time domain gear vibration data with 64 points.

Table 4. Classification results for ARD and PCA on time domain vibration data with 64 points.

| Number of inputs | PCA Classification | ARD Classification |
|---|---|---|
| 3 | 54 | 53 |
| 5 | 55 | 51 |
| 7 | 52 | 47 |
| 10 | 61 | 58 |

#### 4.4.3 Frequency domain input data

From Fig.6 it can be seen that the performance of PCA is slightly better than that of ARD. PCA gives a network classification of 100% at an input dimension of 10 while ARD gives a network performance of 99%. At lower input dimensions PCA is clearly better than ARD. This is due to the fact that ARD discards some of the data, while PCA simply transforms the input data to a smaller dimension.

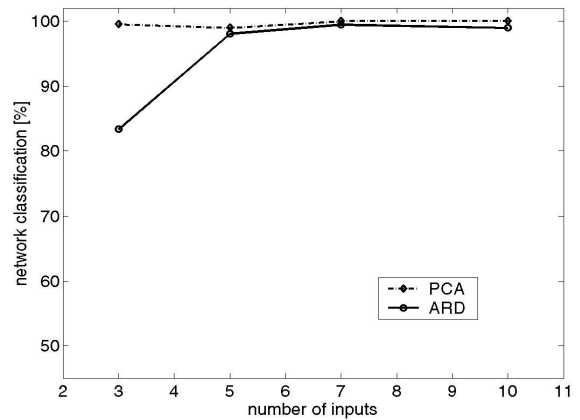

Fig. 6. Plot of network classification results for the frequency domain gear vibration data with 256 points.

Table 5. Classification results for ARD and PCA on frequency domain vibration data with 256 points.

| Number of inputs | PCA Classification | ARD Classification |
|---|---|---|
| 3 | 99.53 | 83.34 |
| 5 | 99.00 | 98.10 |
| 7 | 100.0 | 99.50 |
| 10 | 100.0 | 99.00 |

**4.4.4. Extracted features used as input data**

From Fig.7 it is seen that ARD performs better than PCA. This is an indication that there are some features that are strongly related to the target function. The best network classification is achieved at an input dimension of 7 for both methods. ARD give an average network classification of 98.83% while PCA give a network classification of 91.65 %. This case illustrates the fact that ARD is sometimes better than PCA.

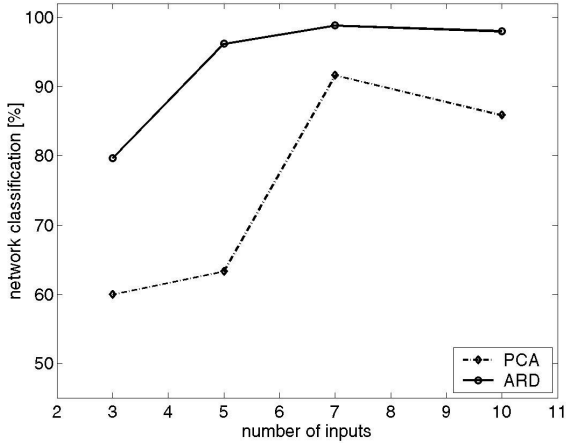

Fig. 7. Plot of network classification results for the frequency domain gear vibration data with 256 points.

Table 6. Classification results for ARD and PCA on the 62 extracted features.

| Number of inputs | PCA Classification | ARD Classification |
|---|---|---|
| 3 | 60.00 | 79.67 |
| 5 | 63.34 | 96.17 |
| 7 | 91.65 | 98.83 |
| 10 | 85.83 | 98.00 |

**5. Conclusion**

Data was pre-processed using PCA and ARD. The inputs to the PCA and ARD are vibration data from a cylinder experiment and time domain vibration data from a gear test rig. The data from PCA and ARD are used to train neural networks for identifying faults in a population of cylinders and gears. It is observed that PCA and ARD can both be used as very effective pre-processing techniques for reducing the input space. PCA performs slightly better than ARD in most of the analyses but there are some cases where PCA cannot be used. In such cases the ARD is a viable option. ARD is computationally more expensive than PCA and therefore PCA should be used whenever possible. ARD and PCA can be used to aid engineering judgement when selecting input features to a neural network. The choice of which method to use is dependent on the data being processed.